\documentclass[preprint,12pt,showkeys,nofootinbib,floatfix]{revtex4}
\usepackage{amssymb}
\usepackage{amsfonts}
\usepackage{amsmath}
\usepackage{graphicx}
\usepackage{subfigure}
\usepackage[dvipdfm,
            pdfstartview=FitH,
            bookmarksnumbered=true,
            bookmarksopen=true,
            colorlinks,
            pdfborder=001,
            linkcolor=green,
            anchorcolor=green,
            citecolor=red
            ]{hyperref}
\usepackage[toc,page,title,titletoc,header]{appendix}
\usepackage{graphics}
\usepackage{epsfig}
\usepackage{slashed}
\usepackage{latexsym}
\usepackage{syntonly}
\newcommand{\be}{\begin{equation}}
\newcommand{\bea}{\begin{eqnarray}}
\newcommand{\eea}{\end{eqnarray}}
\newcommand{\ba}{\begin{array}}
\newcommand{\ea}{\end{array}}
\newcommand{\ee}{\end{equation}}
\def\th{\theta}
\begin{document}
\title{Crystalline geometries from fermionic vortex lattice with hyperscaling violation}
\author{Li-Ke Chen$^{1,2}$}
\author{Hong Guo$^{1}$}
\thanks{Li-Ke Chen and Hong Guo contribute to this paper equivalently.}
\author{Fu-Wen Shu$^{1,2,3}$}
\thanks{E-mail address: shufuwen@ncu.edu.cn}
\affiliation{
$^{1}$Department of Physics, Nanchang University, Nanchang, 330031, P. R. China\\
$^{2}$Center for Relativistic Astrophysics and High Energy Physics, Nanchang University, Nanchang, 330031, P. R. China\\
$^{3}$Kavli Institute for Theoretical Physics China, Institute of Theoretical Physics, Chinese Academy of Sciences,, Beijing, 100190, P. R. China}
\begin{abstract}
We analytically consider the spontaneous formation of a fermionic crystalline geometry in a gravity background with Lifshitz scaling and/or hyperscaling violation. Fermionic vortex lattice solution sourced by the lowest Laundau level has been obtained. Thermodynamic analysis shows that the fermionic vortex lattice favors a triangular configuration, regardless of the values of the Lifshitz scaling $z$ and the hyperscaling violation exponent $\theta$. Our results also show that the larger $z$ or lower $\theta$ leads to more stable lattices thermodynamically.
\end{abstract}

\keywords{Gauge/gravity duality, fermionic vortex lattice, Lifshitz gravity}
\maketitle
\section{Introduction}
Gauge/gravity duality\cite{maldacena,gkp,witten} has attracted a lot of attention in the past few years, especially after the discovery of consistent holographic results, RHIC experiments on the viscosity/entropy-density ratio  \cite{kss,bl,geshu}   and more recent fruitful results of applications of holography to condensed matter physics \cite{hartnoll,herzog,cai1,ling1}. Due to the nature of the duality, it is a promising way of studying gauge theories in the strongly-coupled regime, where the usual perturbative methods fail to apply. Many phenomena in the condensed strongly-coupled systems, such as  the high-$T_c$ superconductivity, the superfluidity and the non-Fermi liquid behavior, have been addressed in the holographic framework. However, most of these works only focus on the systems which are of conformal invariance. Triggered by Son's pioneer work on nonrelativistic holography\cite{son}, recently there has been a great interest in extending gauge/gravity duality to holographic description of QFTs without the conformal invariance. Such extension has been fulfilled to Lorentz-symmetry breaking field theories which exhibit dynamical scaling \cite{klm,taylor} and more recently to theories with hyperscaling violation\cite{kiritsis,xdong}\footnote{For a recent progress, please see Refs. \cite{panzhang,lmx,gv,kym,kofinas,dss,elp,fengg,fl,kpww1,lsachdev,kpww2,br,pdsr} for an incomplete list.}. Nevertheless, all these theories still possess translational and spatial rotational symmetries. As a consequence, many crucial features of the real world materials, say, the effect of the momentum dissipation of charge carries in optical conductivity\cite{horowitz1,horowitz2} , are still far from being achieved.

To build more realistic condensed matter system in the holographic framework, the spatial translational invariance in the bulk must be broken so as to introduce the momentum dissipation. There are several ways to achieve this:

(i) The first way is to induce a holographic lattice by imposing a spatially inhomogeneous periodic source for a scalar field coupled to an Einstein-Maxwell theory \cite{horowitz1}, or alternatively by considering the backreaction of a periodic chemical potential on the metric\cite{horowitz2}.

(ii)The second approach is achieved by introducing a uniform chemical potential into the model\cite{dg1,dg2,donos,dh}. Translational invariance in these models is broken,  but  they have a Bianchi VII$_0$ symmetry, which is associated with the helical order.

(iii) The third one is to treat the massive gravity as a holographic framework of describing theories with broken translational symmetry\cite{vegh,davison,btv,bt}.

(iv) The forth mechanism was proposed in \cite{dg3,aw,kr} where the translational symmetry is broken by introducing massless scalars which lead to a linear dependence on the spatial coordinates of the boundary.

(v) Recently there has a novel approach which was first proposed in \cite{bhks}  and then generalized to the fermionic case\cite{mozaffar} and gravity duals with Lifshitz and/or hyperscaling violation\cite{bh}. The bulk geometry of this model is an $AdS_2 \times R^2$ space supported by a magnetic field, which breaks the translational symmetry. Systems of this type exhibit nonperturbative instabilities of a probe charged scalar field coupled to the magnetic field, and the vortex lattice can be constructed via the instabilities.  Notably,  a distinguished difference between the first mechanism \cite{horowitz1,horowitz2} and this one lies in their behavior at IR. Different from the current case where the effect of the lattice could persist deep in the IR, the first case cannot since the background charge carriers screen the spatially modulated chemical potential in the IR. This feature gives merit to fully understanding the formation of the vortex geometry deep in the IR. Another advantage of this approach is that the backreacted crystalline geometry can be achieved  analytically.

In this paper we adopt the last approach and consider the crystalline geometry in gravity duals with Lifshitz scaling and/or hyperscaling violation, which is  probed by a charged Dirac fermion. We are interested in the spontaneous formation of a fermionic crystalline geometry sourced by the lowest Landau level solutions.  We analytically solve the corresponding coupled PDEs for the metric and the gauge field.  We obtain the same result as  \cite{mozaffar},  which is different from the one obtained in \cite{bhks} where a lattice structure induced by a charged scalar condensate only corrects the background magnetic field. In the fermionic case, however, the backreaction of the fermionic lattice will lead to an emergent electric field and in turn yields an effective charge density.  Furthermore, we also investigate influences of the scaling exponent and the hyperscaling violation exponent on the spontaneous formation of the crystalline geometry. To achieve this, we carefully analyze the thermodynamic quantities of the systems.  Our results show that lattices with larger $z$ or lower $\theta$ are more stable thermodynamically.  Furthermore, it also shows that the fermionic vortex favors a triangular configuration, regardless of the values of $z$ and $\theta$.

The organization of this paper is as follows: In the next section we will discuss the background geometry in question. Equations of motion are obtained and the Dirac field as a probe has been considered and the corresponding Dirac equation in this background has been achieved. In section 3, we solve these differential equations and construct a vortex lattice solution. The radial behavior of the wave functions is also discussed. Some thermodynamical variables such as free energy of the lattice will be discussed in section 4, where we also find that the vortex lattice favors a triangular configuration. Some thermodynamic behavior of the lattice affected by $z$ and $\theta$ has also been discussed there. We give conclusions in the last section. Calculations of double Fourier series are given in the appendix.

\section{Background Setup}\label{0002}
In this section, we will discuss the basic ingredients to build the model of fermionic lattice with hyperscaling violation exponent.
Our start point is the Einstein-Maxwell-Dilaton (EMD) model:
\begin{equation}\label{S1}
S_1 =-\frac{1}{16\pi G}\int
d^{4}x\sqrt{-g}\left[R-\frac{1}{2}(\partial\phi)^2+V(\phi)-\frac{Z(\phi)}{4}F^2\right],
\end{equation}
where the dilaton $\phi$ is dual to the scalar relevant operator of
the system that drives a non-trivial RG flow from the UV to the IR.
It can either be a constant or running in the IR.  The $U(1)$ gauge
field coupled to the dilaton is required to give the anisotropic
scaling. Since we are interested in spontaneous formation of the
crystalline geometries, the translational invariance should be
broken. In this scheme, we achieve this by placing a magnetic filed
along $x$ direction, so the gauge field is restrained as
\begin{eqnarray}
F_{xy}=Q_mdx\wedge dy,
\end{eqnarray}
where $Q_m$ can be viewed as the magnetic charge, and we assume that $Q_m\geq0$ throughout this paper.

We assume that $V(\phi)$ and $Z(\phi)$ have
exponential asymptotics. In particular, we are considering the
following forms
\be
V(\phi)=V_0e^{\gamma\phi},\ \ \ Z(\phi)=e^{\lambda\phi}.
\ee

 The model can be supported by an extremal black
brane whose near-horizon geometry is given by the hyperscaling
violating Lifshitz (hvLif) metric
\begin{eqnarray}\label{metric}
ds^2&=&L^2r^{\theta}\left(-\frac{dt^2}{r^{2z}}+\frac{dr^2}{r^2}+\frac{dx^2+dy^2}{r^2}\right),\\
\label{dilaton}e^{\phi(r)}&=&e^{\phi_0}r^{-\sqrt{\theta^2-2z(\theta-2)-4}},\\
\gamma&=& \frac{\theta}{\sqrt{\theta^2-2z(\theta-2)-4}},\\
\lambda&=& \frac{4-\theta}{\sqrt{\theta^2-2z(\theta-2)-4}},\\
L&=&\frac{Q_me^{\lambda\phi_0}}{\sqrt{2(z-1)(2+z-\theta)}},\\
V_0&=&\frac12(z-1)(1+z-\theta)(2+z-\theta)^2Q_m^2e^{(\lambda-\gamma)\phi_0},
\end{eqnarray}
where  $z$ denotes the Lifshitz scaling, $\theta$ is the
hyperscaling violation exponent, $r\rightarrow 0$ and $r\rightarrow \infty$ correspond to UV and IR respectively. For simplification, we have set $\phi_0=0$ in the following argument.

The NEC imposes constraints on the allowed values for $z$ and
$\theta$\cite{xdong}:
\begin{eqnarray}
&&\text{(i) For}\ \     1\leq z\leq 2:\ \ \th \leq 2(z-1)\ \
\text{or}\ \ 2\leq
\th\leq 2+z,\\
&&\text{(ii) For}\ \    2\leq z\leq 4: \ \ \th\leq 2,\ \ \text{or}\
\
2(z-1)\leq\th\leq 2+z,\\
&&\text{(iii) For}\ \ z>4:\ \  \th\leq 2.
\end{eqnarray}

In order to study the spontaneous formation of a crystalline
geometry sourced by the lowest Landau levels, we consider a massive charged
Dirac field as a probe in the hvLif background
\eqref{metric}\cite{mozaffar}
\begin{eqnarray}\label{S2}
S_{D}=\int{d^4x\sqrt{-g}\,i\,\bar{\Psi} \left(
\frac{1}{2}\left(e_a^{\mu}\Gamma^{a} \overrightarrow{D}_{\mu}-
\overleftarrow{D}_{\mu}e_a^{\mu}\Gamma^{a}\right)-m\right) \Psi},
\end{eqnarray}
where $D_{\mu} \equiv
\partial_{\mu}+\frac{1}{8}\omega_{ab,\mu}[\Gamma^a,\Gamma^b]+iqA_{\mu}$, $\bar{\Psi}=\Psi^{\dag}\Gamma^t$,
while $\omega_{ab,\mu}$ is the spin connection and $e^{\mu}_{a}$ is
the vierbein. Before proceeding, we choose the basis for the Dirac matrices as
\begin{eqnarray}
\Gamma ^t=\left(
\begin{array}{cc}
 -i \sigma ^3 & 0 \\
 0 & -i \sigma ^3
\end{array}
\right),\Gamma ^r=\left(
\begin{array}{cc}
 0 & -i \sigma ^2 \\
 i \sigma ^2 & 0
\end{array}
\right),
\Gamma^x=\left(\begin{array}{cc}
\sigma^2 &0\\
0&-\sigma^2
\end{array}\right),\Gamma^y=\left(
\begin{array}{cc}
 \sigma ^1 & 0 \\
 0 & \sigma ^1
\end{array}
\right),
\end{eqnarray}
and the chiral gamma matrix $\Gamma^5\equiv i\Gamma^r\Gamma^t\Gamma^x\Gamma^y$.

Variation of $S_1+S_D$ leads to the following equations of motion
\begin{eqnarray}\label{motion}
&(\Gamma^{\mu}D_{\mu}-m)\Psi=0,\\
&\nabla_{\mu}\left(e^{\lambda\phi}F^{\mu\nu}\right)=j^{\nu}=q\langle\hat{\bar{\Psi}}\Gamma^{\nu}\hat{\Psi}\rangle,\\
&\nabla^{\mu}(\partial_{\mu}\phi)+V_0\gamma e^{\gamma\phi}=\frac{\lambda e^{\lambda\phi}}{4}F^{\mu\nu}F_{\mu\nu}.\label{motion2}\\
& G_{\mu\nu}=T_{\mu\nu},
\end{eqnarray}
where
\bea
T_{\mu\nu}&=&\frac{1}{4}g_{\mu\nu}\left(V_0e^{\gamma\phi}-\frac12(\partial \phi)^2\right)+\frac14\partial_{\mu}\phi\partial_{\nu}\phi+\frac14e^{\lambda\phi}F_{\mu\rho}F_{\nu}{}^{\rho}-\frac1{16}g_{\mu\nu}e^{\lambda\phi}F^2\nonumber\\
&&-\frac{i}{8}\left[\langle\hat{\bar{\Psi}}e_{a\mu}\Gamma^aD_{\nu}\hat{\Psi}\rangle+h.c.\right]+(\mu\leftrightarrow \nu).
\eea
In deriving the above equations, we have replaced the classical fermionic currents with their quantum mechanical ones, following what Allais, McGreevy and Suh did in their paper\cite{AMS}. The reason is that the zero-temperature fermionic system cannot be treated as the classical gas due to the Pauli's Exclusion Principle. Therefore, as one considers the backreaction of the charged fermions on the holographic geometry, the current generally cannot be treated as the classical one\footnote{ There is another way of dealing with this current. One can treat the fermions as an ideal fluid and use a fermionic equation of state as Hartnoll and his collaborators
did in their electron star papers\cite{hartnoll1,hartnoll2}.}.

In order to impose the boundary conditions, it is convenient to
introduce two sets of projection operators \cite{dtong1}, \bea
P_{\pm}=\frac{1}{2}(1\pm \Gamma^r),\ \ \ \ Q_{(\pm)}=\frac12(1\pm
i\Gamma^x\Gamma^y), \eea under which the spinor can be decomposed
into four components $\Psi_{(\pm)}^{\pm}$. $\Psi^-_{(\pm)}$ are
interpreted as sources, while  $\Psi^+_{(\pm)}$ are responses. The
subscript $(\pm)$ refers to spin-up and spin-down states
respectively.

 Following \cite{mozaffar}, the boundary conditions can be imposed as the following:

 (i) At the UV, we require that $\Psi^-=0$ or $\Psi^+=0$ for standard or alternative quantization respectively. After adding a boundary term to the action such that it has a well-defined variation principle, the UV boundary condition can be translated into the following form
\begin{eqnarray}
\left\{\begin{array}{lcr}
\xi_+=0,\hspace*{0.3cm}\chi_-=0, &\hspace*{1cm} \text{for standard quantization},\\
\xi_-=0,\hspace*{0.3cm}\chi_+=0, &\hspace*{1cm} \text{for alternative quantization}.
\end{array}\right.
\end{eqnarray}
where  $\xi_\pm=\Psi_{(+)}^+\pm \Psi_{(-)}^+$ , $\chi_\pm=\Psi_{(+)}^-\pm \Psi_{(-)}^-$.

(ii) At the IR boundary, as the background metric does not include horizon, it's necessary to consider a cutoff
in the IR boundary if we want to obtain a nontrivial solution for
the fermion in the entire bulk geometry. One possible choice is to impose a hard wall that abruptly cuts the geometry at some finite $r=r_0$. After considering the variation principle, our boundary condition on the hard wall is the same as the one given in \cite{mozaffar}
\begin{eqnarray}
\left\{\begin{array}{lcr}
\tilde{\xi}_-=0,\hspace*{0.3cm}\tilde{\chi}_+=0, &\hspace*{1cm} \text{for standard quantization},\\
\tilde{\xi}_+=0,\hspace*{0.3cm}\tilde{\chi}_-=0, &\hspace*{1cm} \text{for alternative quantization},
\end{array}\right.
\end{eqnarray}
where
\begin{eqnarray}
\left(\begin{array}{c}
\tilde{\xi}_+\\
\tilde{\xi}_-
\end{array}\right)
=\left(\begin{array}{cc}
 \cos\frac{\Omega}{2}& \sin\frac{\Omega}{2}\\
 \sin\frac{\Omega}{2}& -\cos\frac{\Omega}{2}
\end{array}\right) \left(\begin{array}{c}
\Psi^+_{(+)}\\
 \Psi^+_{(-)}
\end{array}\right),\hspace*{0.5cm}
\left(\begin{array}{c}
\tilde{\chi}_+\\
\tilde{\chi}_-
\end{array}\right)
=\left(\begin{array}{cc}
 -\cos\frac{\Omega}{2}& \sin\frac{\Omega}{2}\\
 \sin\frac{\Omega}{2}& \cos\frac{\Omega}{2}
\end{array}\right) \left(\begin{array}{c}
\Psi^-_{(+)}\\
 \Psi^-_{(-)}
\end{array}\right),\nonumber
\end{eqnarray}
where $\Omega \in (-\pi,\pi]$ is a chiral angle.


\section{Vortex lattice solution}
\subsection{Droplet solution}
In this subsection, we will discuss the Fermionic vortex lattice with hyperscaling violation exponent. According to \cite{mozaffar}, the IR instability will lead to a crystalline ground state. So it is convenient to obtain a degenerated $\Psi=0$ solution with a vortex lattice solution.

To proceed, we consider the backreaction of the fermionic field on the background which can be
achieved by expanding the fermionic field and the gauge field around
the critical point,
\begin{eqnarray}\label{psiexp}
\Psi(\textbf{x},r,t)&=&\epsilon\Psi_1(\textbf{x},r,t)+\epsilon^{3}\Psi_3(\textbf{x},r,t)+\ldots\\
\label{Aexp}A_\mu(\textbf{x},r)&=&A^{(0)}_\mu(\textbf{x},r)+\epsilon^2
A^{(2)}_\mu(\textbf{x},r)+\epsilon^4A^{(4)}_{\mu}(\textbf{x},r)\ldots.
\end{eqnarray}
Neglecting the backreaction of $\Psi$ on the gauge sector, as well as rescaling the fermionic field $\Psi_1(r,x,y)=(-h)^{-1/4}\,\psi(r,x,y)$, the Dirac equation in hyperscaling violation metric is
\begin{eqnarray}\label{first}
\left(\Gamma^r \partial_r+\Gamma^x(\partial_x +iqQ_my)+\Gamma^y \partial_y-mLr^{\frac{\theta}{2}-1}\right) \psi=0.
\end{eqnarray}
Acting $(\Gamma^{\mu}D_{\mu}+mLr^{\frac{\theta}{2}-1})$ on the above equation, we obtain a second order differential equation
\begin{eqnarray}\label{second}
\left[\partial_r^2+\partial_x^2+\partial_y^2+2iqQ_{m}y\partial_x-q^2Q_{m}^2y^2+iqQ_m\Gamma^y\Gamma^x-m^2L^2r^{\theta-2}\right. \nonumber\\ \left.-mL\left(\frac{\theta}{2}-1\right)\Gamma^r r^{\frac{\theta}{2}-2}\right]\psi(r,x,y)=0.
\end{eqnarray}

To solve the equation, we notice that $\Gamma^y\Gamma^x$ commutes with $\Gamma^r$. As a result we can expand the field as $\psi^{\pm}(r,x,y)=\rho(r)g(y)e^{ikx}C_{\pm}$ where the simultaneous eigenstates $C_{\pm}$ should satisfy the condition that  $i\Gamma^y\Gamma^x C_{\pm}=\pm C_{\pm}$ and $\Gamma^r C_{\pm}=\pm C_{\pm}$. Therefore the normalized $C_{\pm}$ can be found as
\begin{eqnarray}\label{Cpm}
C_{+}=\frac{\sqrt{2}}{2}\left(
\begin{array}{c}
 0 \\
 1\\
 1\\
 0
\end{array}
\right)
\;\;\;,\;\;\;C_{-}=\frac{\sqrt{2}}{2}\left(
\begin{array}{c}
 1 \\
 0\\
 0\\
 1
\end{array}
\right).
\end{eqnarray}

After doing so, Eq.(\ref{second}) reduces to a set of ODEs
\begin{eqnarray}\label{radial}
\rho_{\pm}''(r)+\left( \lambda_{n_{\pm}}\mp mL\left(\frac{\theta}{2}-1\right)r^{\frac{\theta}{2}-2}-m^2L^2r^{\theta-2}\right)\rho_{\pm}(r)=0\\
g''_{n_{\pm}}(Y)-g_{n_{\pm}}\left(Y^2+\frac{\lambda_{n_{\pm}}}{Q_m}\pm 1\right)=0,\label{veq}
\end{eqnarray}
where $\lambda_{n_{\pm}}$ are constants corresponding to Landau levels and we have introduced a new variable $Y=\sqrt{Q_m}(y+\frac{k}{Q_m})$. An interesting fact is that both Eqs.(\ref{radial}) and (\ref{veq}) are independent of the dynamical exponent $z$, which means that
the exponent $z$ contributes to the vortex only through a prefactor $h$.

The general solution of the above equation is nothing but the familiar Hermite function
\begin{eqnarray}
g_{n_{\pm}}(Y)\sim e^{\frac{-Y^2}{2}}H_{n_{\pm}}(Y),
\end{eqnarray}
the corresponding eigenvalues are given by $\lambda_{n_{\pm}}=-2Q_m(n_{\pm}+\frac{1}{2}\pm\frac{1}{2})$. One thing should be mentioned is that the above Eqs.\eqref{radial} and \eqref{veq} are second order, while the original Dirac equation is a first order one. It may impose a constraint on the eigenvalues $\lambda_{n_{\pm}}$. It turns out in \cite{mozaffar} that this is given by
$\lambda_{n_+}=\lambda_{n_-}$, implying that $n_-=n_++1=n$. In what follows, these relations always hold. Therefore, when we refer to the Landau level it always represents $n$.

\subsection{Vortex solution}
It was found \cite{maeda} that the vortex lattice can be constructed from the droplet solution at the lowest Landau level. One therefore has
\begin{eqnarray}
\psi_0^{lat}=\sum_{l=-\infty}^{\infty}C_l e^{ik_lx}\psi_0(y,k_l),
\end{eqnarray}
where $C_l\equiv e^{-i\pi \frac{a_2}{a_1^2}l^2}$, $k_l=\frac{2\pi
l}{a_1}\sqrt{Q_m}$ and
$\psi_0=e^{-\frac{Q_m}{2}(y+\frac{k_l}{Q_m})^2}$. After using the
alternative elliptic theta function
\begin{eqnarray}
\Theta_3(\nu,\tau)=\sum_{l=-\infty}^{\infty} q^{l^2}Z^{2l},
\end{eqnarray}
 the vortex lattice solution can be rebuilt as
\begin{eqnarray}
\psi_0^{lat}=-e^{-\frac {Q_my^2}{2}}\Theta_3(\upsilon,\tau),
\end{eqnarray}
where $q=e^{i\tau\pi}$, $Z=e^{i\pi\upsilon}$ with $\upsilon=\frac{\sqrt{Q_m}(x+iy)}{a_1}$ and $\tau=\frac{2\pi i-a_2}{a_1^2}$.

A well-known fact is that elliptic theta function has two special properties.  The first one is that $\Theta_3$ has a pseudoperiodicity
\begin{eqnarray}
\Theta_3(\upsilon+1,\tau)&=&\Theta_3(\upsilon,\tau),\\
\Theta_3(\upsilon+\tau,\tau)&=&e^{-2\pi i(\upsilon+\frac{\tau}{2})}\Theta_3(\upsilon,\tau)
\end{eqnarray}
implying that the function $\psi^{lat}_0$ which depends on the form of $\Theta_3$ has the property of the invariance of translation
according to the lattice generators,
\begin{eqnarray}
\overrightarrow{b_1}&=&\frac{1}{\sqrt{Q_m}}a_1\partial_x,\\
\overrightarrow{b_2}&=&\frac{1}{\sqrt{Q_m}}(\frac{2\pi}{a_1}\partial_y+\frac{a_2}{a_1}\partial_x).
\end{eqnarray}

The second property is that the $\Theta_3$ function will vanish at
 \begin{eqnarray}
\overrightarrow{\chi}_{m,n}=(m+\frac{1}{2})\overrightarrow{b_1}+(n+\frac{1}{2})\overrightarrow{b_2}&&m,n\in\mathbb{N}
\end{eqnarray}
and the phase of $\langle\mathcal{O}\rangle$ rotates by $2\pi$. So the core of vortex are located at $\overrightarrow{\chi}_{m,n}$.

\subsection{Radial equation}
Since we are focusing on the lowest Landau level solutions, we have $\lambda_+=\lambda_-=0$. Substituting this into Eq.\eqref{radial}, we get
\begin{eqnarray}\label{radial2}
\rho_{\pm}''+\left[\mp mL\left(\frac{\theta}{2}-1\right)r^{\frac{\theta}{2}-2}-m^2L^2r^{\theta-2}\right]\rho_{\pm}(r)=0.
\end{eqnarray}
Notice that equations for $\rho_+$ and $\rho_-$ are related through
a transformation $mL\rightarrow -mL$. Therefore one can always
obtain $\rho_-$ from $\rho_+$ by inversing the value of $mL$, and
vice versa. In what follows, we only consider the equation for
$\rho_+$, and we omit the subscript as well,
\begin{eqnarray}\label{radial3}
\rho''+\left[mL\left(\frac{\theta}{2}-1\right)r^{\frac{\theta}{2}-2}+m^2L^2r^{\theta-2}\right]\rho(r)=0.
\end{eqnarray}
\subsubsection{$\theta=0$}
This case reduces to a Lifshitz spacetimes without hyperscaling violation. The radial equation \eqref{radial3} becomes
\begin{eqnarray}
\frac{\rho''(r)}{\rho(r)}+mLr^{-2}-m^2L^2r^{-2}= 0,
\end{eqnarray}
The above equation admits a power-law solution
\begin{eqnarray}
\rho=c_1r^{\alpha_+}+c_2r^{\alpha_-},
\end{eqnarray}
where
$$\alpha_{\pm}=\pm\left(mL-\frac{1}{2}\right).
$$

Translating it into the fermionic field, we obtain
\be
\Psi_1(r,x,y)=\frac1{L^{3/2}}\left(c_1r^{\Delta_+}+c_2r^{\Delta_-}\right)\psi_0^{lat}(x,y)C_{+}.
\ee
where $\Delta_{\pm}=1+\frac{z}2+\alpha_{\pm}$.
The above solution has two radial modes. For the case where $mL>\frac{3+z}{2}$ one can only consider the standard quantization\footnote{This corresponds to $\Delta_-<0$. Since we are considering spontaneous formation of the lattice, we should turn off the source term and leave a scaling dimension of the boundary operator to be $\Delta(\Psi_1)=\Delta_+$.}, while for $mL<\frac{3+z}{2}$ the alternative quantization is also available.

\subsubsection{$\theta\neq 0$}
For this case, we first define a function
$\varphi(r)\equiv\frac{\rho'(r)}{\rho(r)}$, then the equation
\eqref{radial3} becomes a Ricatti equation
\begin{eqnarray}\label{F}
 \varphi'(r)+\varphi^2(r)-[(mLr^{\frac{\theta}{2}-1})'+(mLr^{\frac{\theta}{2}-1})^2]=0,
\end{eqnarray}
which admits a special power-law solution
\begin{eqnarray}\label{ssol}
\varphi_0(r)=mLr^{\frac{\theta}{2}-1}.
\end{eqnarray}
We therefore get an exponential form of $\rho$
\be\label{rho_0} \rho(r)=\rho_0
\exp{\left(\frac{2mL}{\theta}r^{\frac{\theta}{2}}\right)}, \ee
where $\rho_0=\rho(r=0)$ is a constant.

Actually, making use of the special solution $\varphi_0(r)$
(\ref{ssol}), more general solutions of Eq.(\ref{F}) can be found with
the assumption $\varphi(r)=u(r)+\varphi_0(r)$. Substituting this
into \eqref{F}, one get a solution for $u(r)$
\begin{eqnarray}
u(r)=\frac{\lambda\theta \exp{\left(-\left(\lambda r\right)^{\frac{\theta}{2}}\right)}}{b\lambda\theta-2\Gamma\left(\frac{2}{\theta},\left(\lambda r\right)^{\frac{\theta}{2}}\right)},
\end{eqnarray}
where $b$ is an integration constant and
\be
\lambda\equiv \left(\frac{4mL}{\theta}\right)^{\frac{2}{\theta}}.
\ee
We therefore obtain a general solution of $\rho$ by integrating
\begin{eqnarray}
\rho(r)=\rho_0\exp\left(\frac{2mL}{\theta}r^{\frac{\theta}{2}}+\int u(r)dr \right).
\end{eqnarray}
Near UV boundary ($r\rightarrow 0$), the above function for $\theta>0$ can be expanded as
\be
\rho(r )=d_1+d_2 r+\mathcal{O}(r )
\ee
where
\be
d_1=\rho_0\left(1+\frac{\theta}2\right)\left(\frac{b\lambda \theta}{2}-\Gamma\left(\frac2{\theta}\right)\right),\ \ \ d_2=\frac{\rho_0}{2}\left(1+\frac{\theta}{2}\right)\theta.
\ee
Translating it into the fermionic field, we obtain
\be
\Psi_1(r,x,y)=\frac1{L^{3/2}}\left(d_1r^{\Delta_-}+d_2r^{\Delta_+}\right)\psi_0^{lat}(x,y)C_{+},
\ee
where $\Delta_{\pm}=\frac{6+2z-3\theta}{4}\pm\frac12$. For the case where $ 3\theta-2z>4$ one can only consider the standard quantization, and we turn off the source term, leaving $\Delta(\Psi_1)=\Delta_+$. From now on, we pay our attention to this case and for simplification we briefly denote $\Delta_+$ as $\Delta$.

\section{the free energy}
\subsection{Linearized backreaction on the metric}
In this subsection, the backreactions of the fermionic vortex on the metric, the gauge field and dilaton will be discussed. The backreactions are sourced by the fermions at matter current and energy-momentum tensor at order $\mathcal{O}(\epsilon^2)$. It shows that the only nontrivial sources at order $\mathcal{O}(\epsilon^2)$ are $T_{tx}$ and $T_{ty}$. So that the following ansatz for the backreacted metric, gauge field and dilaton, respectively, are:
\begin{eqnarray}\label{ds2}
&&ds^2=L^2r^{\theta}(-\frac{dt^2}{r^{2z}}+\frac{dr^2}{r^2}+\frac{dx^2+dy^2}{r^2})+L^2\epsilon^2r^\beta[a(r,x,y)dtdx+b(r,x,y)dtdy]\\
&&A=Q_m y dx + \epsilon^2r^{\alpha}a^t_2(r,x,y)dt \label{A2} \\
&&\phi(r,x,y)=\phi_1(r)+\epsilon^2\phi_2(r,x,y) \label{phi2}
\end{eqnarray}
where $\beta=\frac{3}{2}\theta-z$,$\alpha=\frac{\theta}{2}-z+4$ and $\phi_1(r)$ is given in \eqref{dilaton}.

At order $\mathcal{O}(\epsilon^2)$, the nontrivial Einstein equations (coming from $G_{tr},G_{tx},G_{ty}$), gauge field equation and dilation equation are shown as (We have set $Q_m=1$ in the following argument)
\begin{eqnarray}\label{twoorder}
&&(\beta+2z+2\Delta-\theta)(\partial_x a+\partial_y b)=0,\\
&&M a=-2i[\langle\hat{\Psi}_1\partial_x\hat{\Psi}^{\dag}_1\rangle-\langle\hat{\Psi}^{\dag}_1\partial_x\hat{\Psi}_1\rangle-2i y \langle\hat{\Psi}^{\dag}_1\hat{\Psi}_1\rangle],\\
&&M b=-2i[\langle\hat{\Psi}_0\partial_y\hat{\Psi}^{\dag}_1\rangle-\langle\hat{\Psi}^{\dag}_1\partial_y\hat{\Psi}_1\rangle],\\
&&Qa^t_2+\frac{1}{2}(\partial_x b-\partial_y a)=2L^3\langle\hat{\Psi}^{\dag}_1\hat{\Psi}_1\rangle,\\
&&r^2\partial^2_r \phi_2+r(\theta-z-1)\partial_r \phi_2+P \phi_2=0\label{twoorder2},
\end{eqnarray}
where
\begin{eqnarray}
&&M=\frac{\sqrt{(z-1)(2+z-\theta)}(-4-10z^2+16\Delta(\Delta+\theta)+\theta(\theta+6+6z)-z(8\Delta+6))}{\sqrt{2}},\\
&&Q=-\frac{1}{4}(2z-4\Delta-\theta-8)(4\Delta+3\theta-2),\\
&&P=\frac{(2+z-\theta)\left(-32+16\theta-3\theta^2+\theta^3+z(32-16\theta+\theta^2)\right)}{2(2z-\theta-2)(\theta-2)}.
\end{eqnarray}

The above equations have several nontrivial features. First of all, equation governing $\phi_2$ is completely decoupled from other variables, which means that the second-order corrections of dilation, if exists, are neither sourced by the vortex lattices directly nor affected through other fields like $a, b$  and $a_2^t$ indirectly. In other words, the vortex lattice does NOT impose any influences on the dilaton at this order. This reminds us that $\phi_2=0$ is a reasonable solution to \eqref{twoorder2}.   Secondly, all the functions except $\phi_2(r,x,y)$ are source by $ \langle\hat{\Psi}^{\dag}_1\hat{\Psi}_1\rangle$ which scale as $r^{2\Delta}$ near the boundary. It is very natural to assume that $a(r,x,y),b(r,x,y)$ and $a^t_2(r,x,y)$ all scale as $r^{2\Delta}$ where $\Delta$ is given by
\begin{eqnarray}
\Delta=
\left\{\begin{array}{lcr}
mL+\frac{1+z}{2},\hspace*{0.3cm} &\hspace*{1cm} \text{for $\theta=0$},\\
2+\frac{z}{2}-\frac{3\theta}{4},\hspace*{0.3cm} &\hspace*{1cm} \text{for $\theta> 0$}.
\end{array}\right.
\end{eqnarray}
as what have shown in the last section. As a consequence we suppose that
\begin{equation}
f_i(r,x,y)=r^{2\Delta}f_i(x,y)
\end{equation}
where $f_i(r,x,y)=a,b,a^t_2$.

Considering that the vortex lattice is periodic in both $x$ and $y$ directions with periodicity $|\vec{b}_1|$ in $x$ direction and $|\vec{b}_2|$ in $y$ direction, we therefore can perform the following double Fourier series:
\begin{equation}\label{dfs}
f_i(x,y)=\frac{1}{L^{3/2}}\sum_{k,l,j} a_1 e^{\frac{2\pi ikx}{|\mathbf{\overrightarrow{b}_1}|}}e^{\frac{2\pi ijy}{|\mathbf{\overrightarrow{b}_2|}}}e^{g(k,l,j)}\widetilde{f_i}(k,l,j),
\end{equation}
where $g(k,l,j)=-\frac{\pi^2k^2}{a^2_1}-i\pi\frac{a_2}{a^2_1}(2l-k)k-i\pi kj-\frac{a^2_1 j^2}{4}$. After inserting above series into Eqs.(\ref{twoorder})-\eqref{twoorder2}, we obtain a set of algebraic equations for $\widetilde{f_i}(k,l,j)$. A trick here is that to avoid doing double Fourier series of derivatives   appearing in the equations, one should first translate them into functions of $\Psi_1$ or $\Psi^{\dag}_1$ and their derivatives (one can find detailed calculations in the appendix) . In the end we  get the following solutions
\begin{eqnarray}
&&\widetilde{a}=\frac{ija_1}{\sqrt{\pi}M},\\
&&\widetilde{b}=\frac{-2i\sqrt{\pi}k}{Ma_1},\\
&&\widetilde{a^t_2}=W=\frac{1}{\sqrt{\pi}Q}(L^3+\frac{2 a_1^2 j^2}{M}-\frac{8\pi^2 k^2}{a_1^2 M}),\\
&&\phi_2=0.
\end{eqnarray}
Substituting above solutions into \eqref{dfs} one obtains solutions in coordinate space.

\subsection{Free energy}
\begin{figure}[htbp]
 \begin{minipage}{1\hsize}
\begin{center}
\includegraphics[scale=0.5]{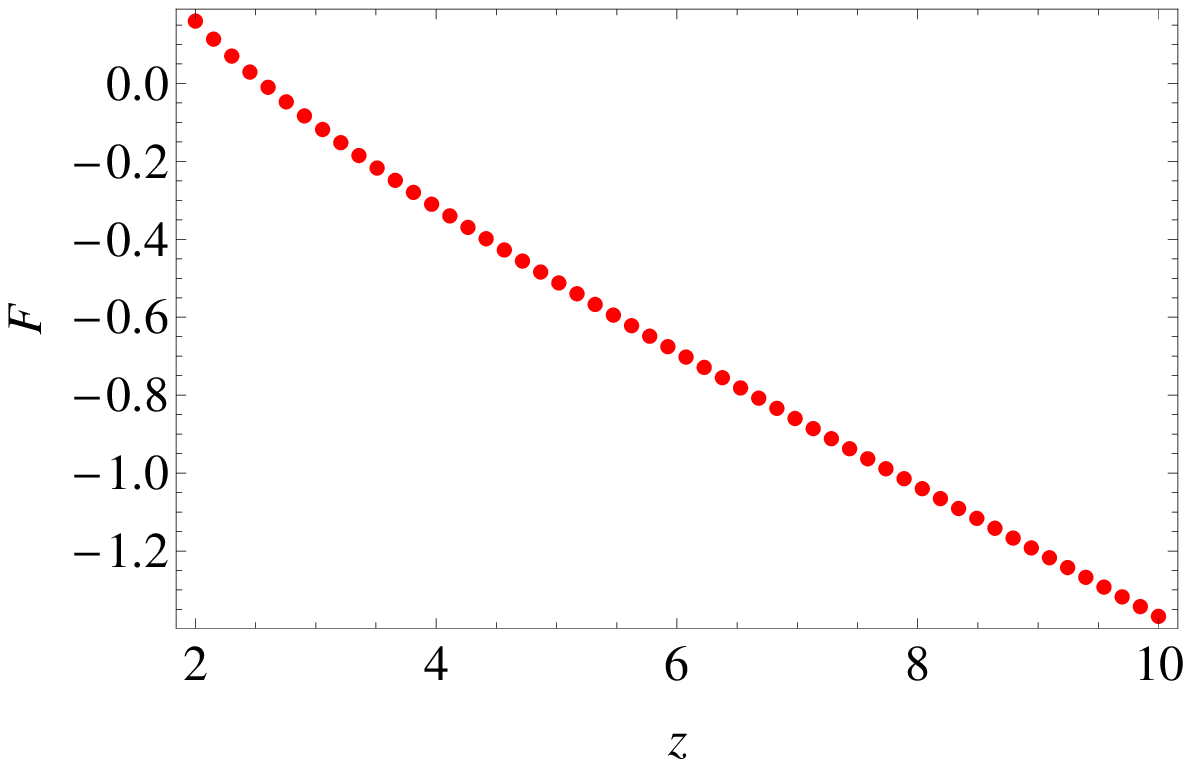}
\includegraphics[scale=0.5]{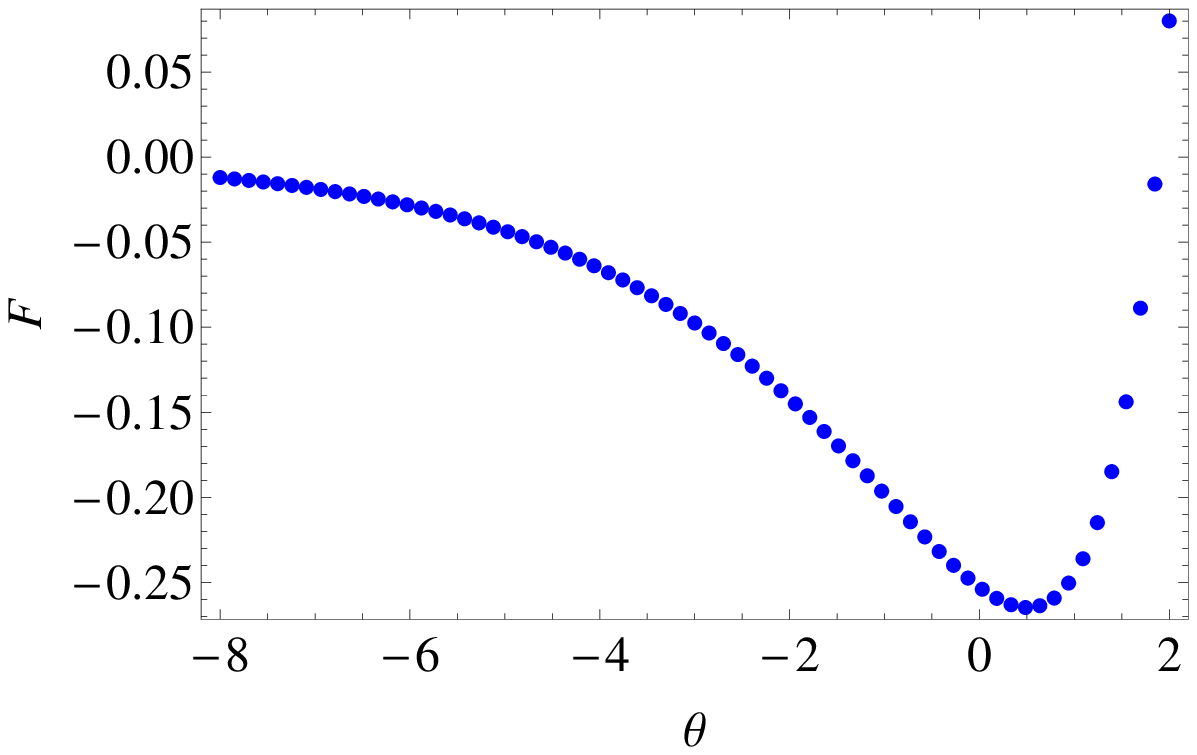}
\end{center}
\caption{ \label{zero} \emph{The zeroth-order free energy is plotted as a function of the Lifshitz scaling $z$ (left plot) and hyperscaling violation exponent $\theta$ (right plot). It is decreasing (almost linearly) with increasing $z$ as shown in the left plot and  increases with increasing $\theta$ as shown in the right plot. }}
\end{minipage}
\end{figure}

\begin{figure}[htbp]
 \begin{minipage}{1\hsize}
\begin{center}
\includegraphics[scale=0.5]{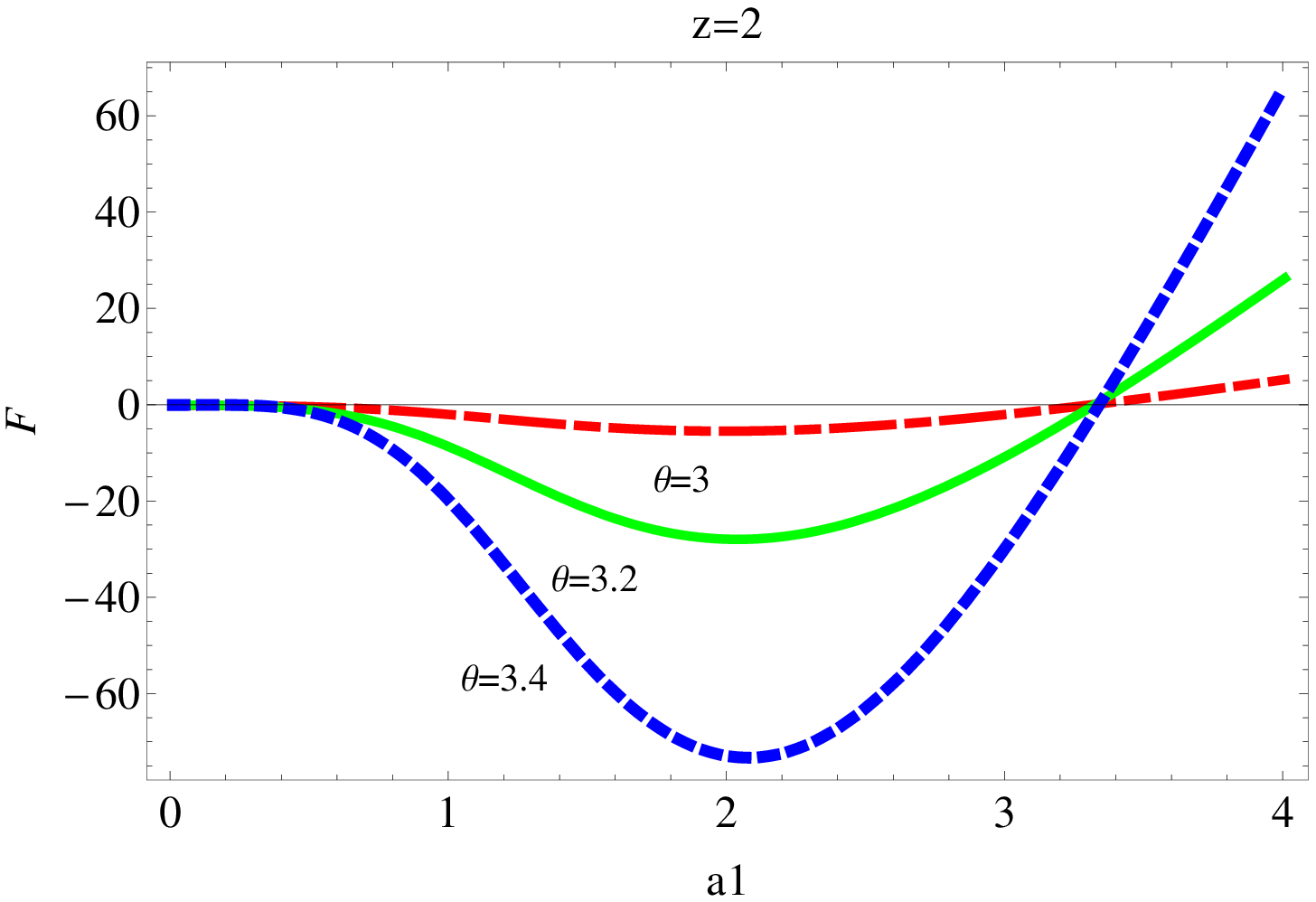}
\includegraphics[scale=0.5]{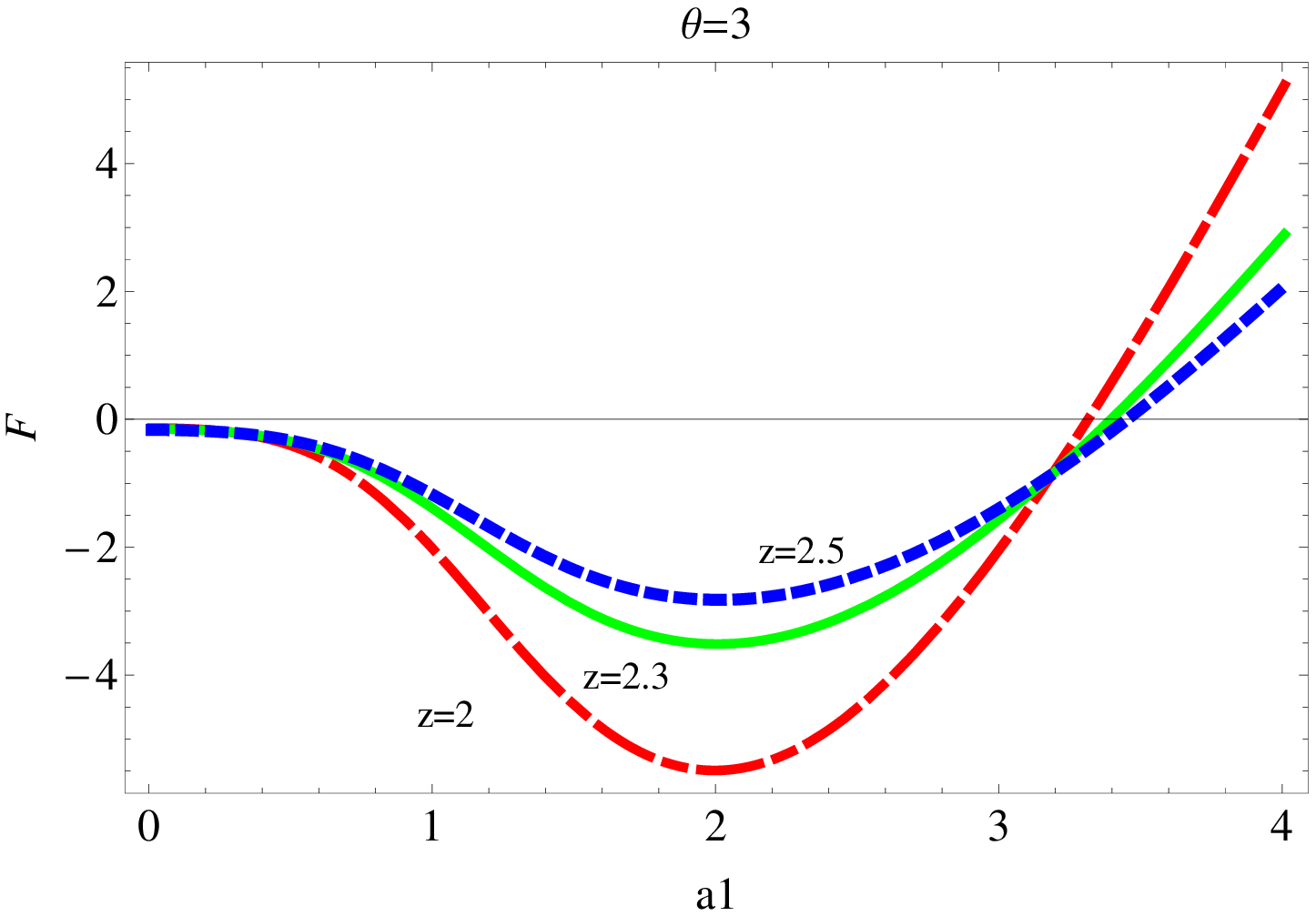}
\end{center}
\caption{ \label{forth} \emph{The forth-order free energy is plotted as a function of $a_1$ with $a_2=\frac{a_1^2}{2}$. The left plot corresponds to curves of forth-order free energy v.s. $a_1$ for different $\theta$, while the right one refers to the corresponding curves for different $z$. Both plots show a minimum located at the neighbor of $a_1=2$.}}
\end{minipage}
\end{figure}

In order to see which configuration the vortex lattice prefers to form, in this subsection we would like to discuss the effects of the lattice formation on the thermodynamic functions. Particularly, we will compute corrections to the free energy of the vortex lattice solution via the on-shell value of the bulk action
\begin{equation}
F\sim T S_{on-shell}
\end{equation}
where $T$ is the energy scale. Following \cite{bhks,bh} we interpret $r_0$ as a confinement scale $\Lambda^{-1}$ in the wall geometry. In this way we substitute $r_0\sim T^{-\frac{1}{z}}$ so as to get temperature dependence. Direct calculations show that:
\begin{eqnarray}
&S_{on-shell}^{(0)}&=\frac{2+5z+z^2-3\theta-3z\theta}{\theta-z-2}r^{\theta-z-2},\\
&S_{on-shell}^{(2)}&=0,\\
&S_{on-shell}^{(4)}&=\frac{1}{16M\pi}\sum_{k,l,j=-\infty}^{\infty}e^{\frac{4\pi ikx}{|\vec{b}_1|}+\frac{4\pi ijy}{|\vec{b}_2|}+2g(k,j,l)}\left(A_1 +A_2r^{2}\right)r^{8+z-\theta},
\end{eqnarray}
where
\begin{eqnarray}
A_1&=&-\frac{4L^2\pi^2k^2+L^2b_1^4j^2}{8+z-\theta}\left(56\Delta^2+4z^2+5\theta^2+14\beta^2-16\theta\beta-32\Delta\theta-10z\theta+56\Delta\beta\right.\nonumber \\
&&\left.+16z\beta+32z\Delta-8\theta+8\beta+8z+16\Delta+1+(\theta-2)(\theta-2z+2) -\frac{4}{L^2}-2L^2V_0\right),\\
\nonumber A_2&=&\left\{56L^2\pi^2\left(\frac{4\pi^2k^4}{b_1^2}+\frac{b_1^4j^4}{b_2^2}+\frac{4\pi b_1 j^2k^2}{b_2}\right)-32\pi M^2 W^2\left[(\alpha+2\Delta)^2-4\pi^2\left(\frac{k^2}{b_2^2}+\frac{j^2}{b_2^2}\right)\right]\right.\\
&&\left.-64MW\pi^{3/2}\left(\frac{b_1^2j^2}{b_2}+\frac{2\pi k^2}{b_1}\right)\right\}\frac1{10+z-\theta}.
\end{eqnarray}
with $b_1=|\vec{b}_1|=a_1, b_2=|\vec{b}_2|=\sqrt{a_2^2+4\pi^2}/{a_1}$.
As a result, up to order $\epsilon^4$, the on-shell action is
\begin{eqnarray}
S_{on-shell}\sim r^{\theta-z-2}_0\left[1+\epsilon^4\left(A_1r_0^{10+2z-2\theta}+A_2r_0^{12+2z-2\theta}\right) +\ldots\right]
\end{eqnarray}
The free energy is then given by
\begin{eqnarray}
F\sim T^{2+\frac{2-\theta}{z}}\left[1+\epsilon^4\left(A_1T^{\frac{-1}{z}(10+2z-2\theta)}+A_2T^{\frac{-1}{z}(12+2z-2\theta)}\right)+\ldots\right]
\end{eqnarray}
Several remarkable remarks are as follows:\\
(i) There is vanishing corrections of free energy at the second order, This is different from the $AdS_2\times R^2$ case as shown both in fermionic crystalline geometry \cite{mozaffar} and in scalar crystalline geometry \cite{bhks}, and is different even from the case where the background spacetime is a  Lifshitz one with  hyperscaling violation \cite{bh}.\\
(ii) Straightforward calculations show that  the free energy decreases, almost linearly, with $z$ for fixed hyperscaling violation exponent $\theta$, while   it increases with $\theta$ for some fixed $z$. This indicates that the vortex lattice has more stable thermodynamic stability for larger $z$ or lower $\theta$. The detailed behavior can be found in  Fig.(\ref{zero}).\\
(iii) More interesting features appear at the forth order where one can show that the vortex lattice favors a triangular configuration, regardless of the values of $z$ and $\theta$. To learn this more explicitly, a plot of free energy v.s. lattice constant $a_1$ ($a_2=\frac12 a_1^2$) has been drawn (the Fig.(\ref{forth})). From those figures, we see that  the free energy has a minimum value for different  $z$ and $\theta$. Remarkably, all the minimums are located very closely to $a_1=2$. It is well known that the equilateral triangular lattice has a lattice constant
$$
a_1=\frac{2\sqrt{\pi}}{3^{\frac14}}\simeq 2.69.
$$
There is a discrepancy between this value and our minimum location $a_1=2$. This discrepancy  possibly comes from a fact that our on-shell action at the forth order is not complete. The complete forth-order on-shell action includes not only the quadratic terms of the second-order ones like $a,b, a_2^t$, but also those terms that are further backreacted by these backreactions $a,b$ and $a_2^t$. In the present paper we have only  considered the first contributions. The second one refers to solving higher-order differential equations for fields like $\Psi_3(r,x,y)$ and $A^{(4)}_{\mu}(r,x,y)$ in Eqs.\eqref{psiexp} and \eqref{Aexp}  and therefore has not yet been taken into account. Although it is complicated, it is possible to do that, following what we have done in a recent paper on vortex lattice formation of a $d-$wave superconductor \cite{shu}. We leave this calculation for future work.

(iv) One less important observation in Fig.\eqref{forth} is that it seems that larger $z$ or $\theta$ has larger minimum near $a_1=2$. This implies that trends to form a triangular lattice decrease with the increasing $z$ and $\theta$.

\begin{figure}[htbp]
 \begin{minipage}{1\hsize}
\begin{center}
\includegraphics[scale=0.5]{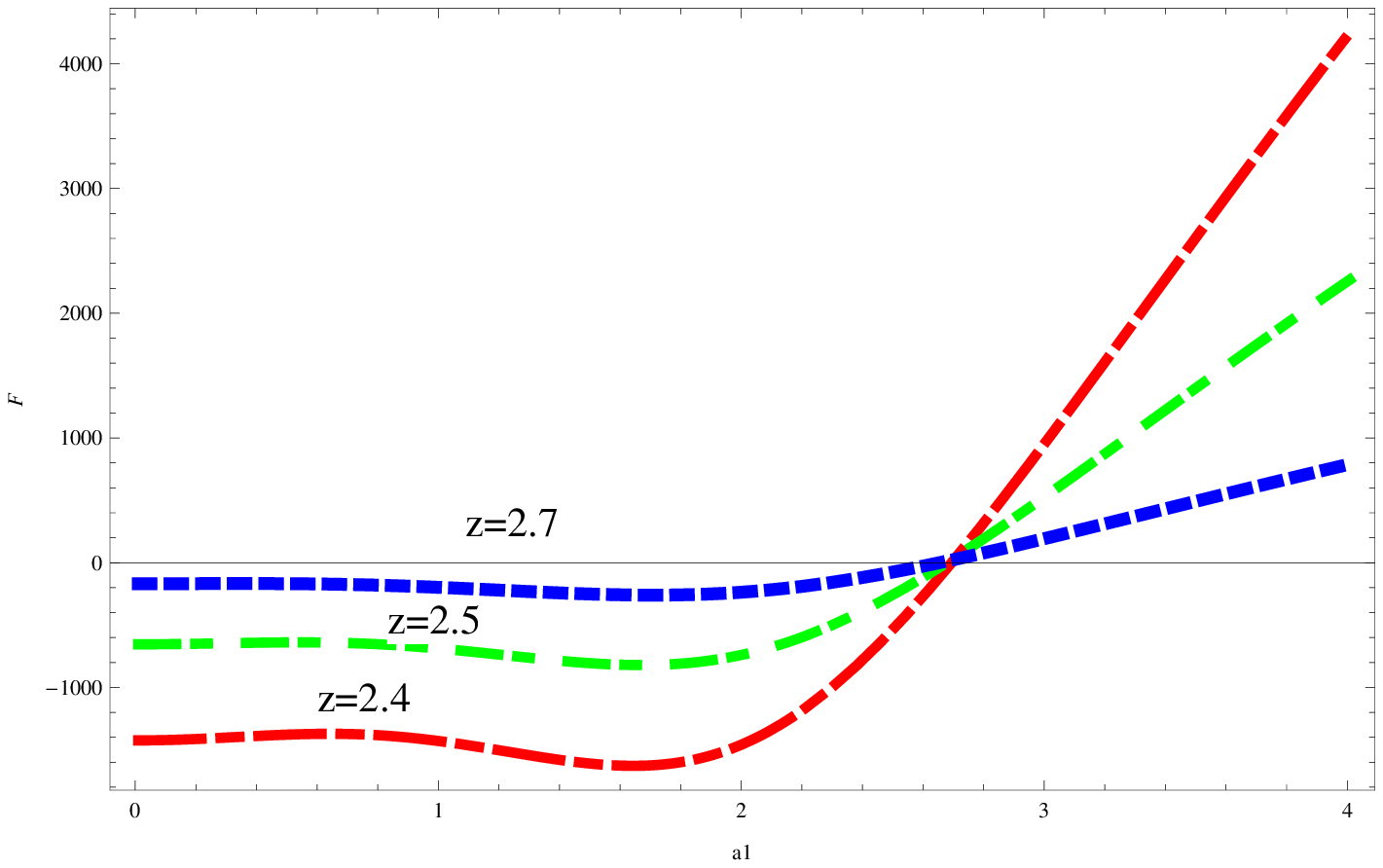}
\end{center}
\caption{ \label{theta0} \emph{The forth-order free energy is plotted as a function of $a_1$ with $a_2=\frac{a_1^2}{2}$ for different $z$, but with fixed exponent $\theta=0$.}}
\end{minipage}
\end{figure}

(v) For a special case $\theta=0$, we have the similar behavior as the nonvaninshing ones, as plotted in Fig.\eqref{theta0}.
\section{Conclusions}
In this paper we have considered the spontaneous formation of a fermionic crystalline geometry in bulk geometry with Lifshitz scaling and/or hyperscaling violation. Fermionic vortex lattice solution sourced by the lowest Landau level has been obtained. The main results of this work are as follows:

(i) The same result as \cite{mozaffar}  has been obtained, that is, different from the one in  \cite{bhks} where a lattice structure induced by a charged scalar condensate only corrects the background magnetic field, in our case the backreaction of the fermionic lattice will lead to an emergent electric field and an effective charge density.

(ii) Contrary to the $AdS_2\times R^2$ case in \cite{mozaffar} and  \cite{bhks}, and the hyperscaling violation case in \cite{bh}, the free energy in our case receives vanishing corrections at the second order.

 (iii) Influences of the scaling exponent and the hyperscaling violation exponent to the spontaneous formation of the crystalline geometry are shown in Figure \eqref{zero}. It can be seen that larger $z$ or lower $\theta$ leads to more thermodynamically stable lattices.

(iv) Our calculation on the free energy shows that the fermionic vortex favors a triangular configuration, regardless of the values of $z$ and $\theta$.

One important point is that in this work we have only considered the vortex solution sourced by the lowest Landau level where the fermionic current can be reduced to a classical one as shown in the appendix \eqref{A8}. However, it is possible to generalize our results to any Landau level where the fermionic current cannot be treated classically and should be replaced by \eqref{A5} and \eqref{A6}. The corresponding free energy becomes
\begin{eqnarray}
F\sim T^{2+\frac{2-\theta}{z}}\sum_{n}\left[\theta(-\lambda_n)+\theta(-\lambda_n)^2\epsilon^4\left(A_n T^{\frac{-1}{z}(10+2z-2\theta)}+B_n T^{\frac{-1}{z}(12+2z-2\theta)}\right)+\ldots\right],
\end{eqnarray}
where $A_n$ and $B_n$ are some constants.

Further investigations and generalizations of this work are possible. As mentioned before, it is of interest to study the full forth-order on-shell action and see if the discrepancy of the lattice constant between our favorite value  (i.e., $a_1\simeq 2$) and the equilateral triangular one where $a_1=2.69$ would be removed. Moreover, it is also interesting to study the formation of the crystalline geometries for spacetimes with black brane horizon. In addition, it is also possible to consider the crystalline geometry in the framework of modified gravity, such as the Ho\v{r}ava-Lifshitz (HL) gravity \cite{HL} proposed recently by Ho\v{r}ava. Indeed, it was found that HL gravity is a minimal holographic dual for the field with Lifshitz scaling \cite{HL2}. Our recent works \cite{shu1,shu2} found that various Lifshitz spacetimes are possible even without matter fields. It is of particular interests to see how to construct the crystalline geometry in this framework.

\section*{\bf Acknowledgements}

This work was supported in part by the National Natural Science Foundation of China (under Grant Nos. 11465012 and 11005165), the Natural Science Foundation of Jiangxi Province (under Grant No. 20142BAB202007) and the 555 talent project of Jiangxi Province.

\appendix

\section{Double Fourier series}
In this appendix we list several key procedures in doing the double Fourier series. Firstly, we expand
\begin{eqnarray}
\hat{\Psi}^{\dag}_{1}\hat{\Psi}_{1}&\sim& \frac{1}{L^{3/2}}\sum_l e^{-i\pi\frac{a_2 l^2}{a_1^2}}e^{\frac{2\pi il}{|\mathbf{\overrightarrow{b}}_1|}x}e^{-\frac{1}{2}(y+\frac{2\pi l}{a_1})^2}
\cdot \sum_{\widetilde{k}=l-k}e^{i\pi\frac{a_2 \widetilde{k}^2}{a_1^2}}e^{-\frac{2\pi i\widetilde{k}}{|\mathbf{\overrightarrow{b}}_1|}x}e^{-\frac{1}{2}(y+\frac{2\pi \widetilde{k}}{a_1})^2}\nonumber \\
&=&\frac{1}{L^{3/2}}\sum_{k,l}e^{-i\pi\frac{a_2}{a_1^2}[l^2-(l-k)^2]}e^{\frac{2\pi ik}{|\mathbf{\overrightarrow{b}}_1|}x}h(y)
\end{eqnarray}
where $h(y)$ can be described as $\sum_j h(j)e^{\frac{2\pi ijy}{|\mathbf{\overrightarrow{b}}_2|}}$ which
\begin{eqnarray}
h(j)&=&\frac{1}{|\mathbf{\overrightarrow{b}}_2|}\int dy h\left(\frac{y}{|\mathbf{\overrightarrow{b}}_2|}\frac{2\pi}{a_1}\right)e^{-\frac{2\pi ijy}{|\mathbf{\overrightarrow{b}}_2|}}\nonumber  \\
&=& \frac{a_1}{2\sqrt{\pi}}e^{A(k,l,j)},
\end{eqnarray}
where
$$
A(k,l,j)=-\frac{\pi^2k^2}{a_1^2}-i\pi kj-\frac{a_1^2j^2}{4}.
$$
We therefore have
\begin{eqnarray}
\hat{\Psi}^{\dag}_{1}\hat{\Psi}_{1}&\sim& \frac{1}{L^{3/2}}\sum_{k,l,j}\frac{a_1}{2\sqrt{\pi}}e^{\frac{2\pi ik}{|\mathbf{\overrightarrow{b}}_1|}x}e^{\frac{2\pi ij}{|\mathbf{\overrightarrow{b}}_2|}y}
e^{g(k,l,j)}\equiv\sum_{k,l,j}\Psi_{k,l,j}^{\dag}\Psi_{k,l,j}
\end{eqnarray}
where
\begin{eqnarray}
g(k,l,j)&=&A(k,l,j)-i\pi\frac{a_2}{a_1^2}[l^2-(l-k)^2]\nonumber \\&=& -\frac{\pi^2k^2}{a^2_1}-i\pi\frac{a_2}{a_1^2}(2l-k)k-i\pi kj-\frac{a^2_1 j^2}{4}.
\end{eqnarray}

According to \cite{AMS}, we have
\begin{eqnarray}\label{A5}
\langle\hat{\Psi}^{\dag}_{1}\hat{\Psi}_{1}\rangle=\Delta n_{n,k,l,j}=n_{n,k,l,j}|_{Q_m}-n_{n,k,l,j}|_{Q_m=0}=n_{n,k,l,j},
\end{eqnarray}
where
\begin{eqnarray}\label{A6}
n_{n,k,l,j}=\sum_{n,k,l,j}\theta(-\lambda_{n})\Psi_{k,l,j}^{\dag}\Psi_{k,l,j},
\end{eqnarray}
and
\begin{eqnarray}
\theta(x)=
 \begin{cases}
 0, &\text{$x<0$},\\
1, &\text{$x\geq 0$},
\end{cases}
 \end{eqnarray}is the step function and $\lambda_n$ is the Landau level. For the lowest Landau level $\lambda_0=0$ we have
\begin{eqnarray}\label{A8}
\langle\hat{\Psi}^{\dag}_{1}\hat{\Psi}_{1}\rangle=\Delta n_{n,k,l,j}=\sum_{n,k,l,j}\Psi_{k,l,j}^{\dag}\Psi_{k,l,j}=\hat{\Psi}^{\dag}_{1}\hat{\Psi}_{1}.
\end{eqnarray}

And we get that
\begin{eqnarray}
\langle\hat{\Psi}_1\partial_y\hat{\Psi}_1^{\dag}\rangle-\langle\hat{\Psi}_1^{\dag}\partial_y\hat{\Psi}_1\rangle&=&\frac{2\pi(-l+k+l)}{a_1}\langle\hat{\Psi}^{\dag}_{1}\hat{\Psi}_{1}\rangle\nonumber \\&=&\frac{2\pi k}{a_1}\langle\hat{\Psi}^{\dag}_{1}\hat{\Psi}_{1}\rangle,\\
\langle\hat{\Psi}_1\partial_x\hat{\Psi}_1^{\dag}\rangle-\langle\hat{\Psi}_1^{\dag}\partial_x\hat{\Psi}_1\rangle-2iy\langle\hat{\Psi}_1^{\dag}\hat{\Psi}_1\rangle&=&\{\frac{2\pi i(-2l+k)}{a_1}+\frac{4i\pi}{a_1}[(l-\frac{k}{2})+\frac{ia_1^2j}{4\pi}]\}\langle\hat{\Psi}^{\dag}_{1}\hat{\Psi}_{1}\rangle
\nonumber \\&=&-a_1 j\langle\hat{\Psi}^{\dag}_{1}\hat{\Psi}_{1}\rangle,
\end{eqnarray}
\begin{eqnarray}
\partial_x a&=&\frac{2i}{M}\langle\hat{\Psi}_1\partial_x^2\hat{\Psi}_1^{\dag}\rangle-\langle\hat{\Psi}_1^{\dag}\partial_x^2\hat{\Psi}_1\rangle-
2iy[\langle\hat{\Psi}_1\partial_x\hat{\Psi}_1^{\dag}\rangle+\langle\hat{\Psi}_1^{\dag}\partial_x\hat{\Psi}_1\rangle]
\nonumber \\ &=& \frac{2\sqrt{\pi}k}{M}ja_1\langle\hat{\Psi}^{\dag}_{1}\hat{\Psi}_{1}\rangle ,\\
\partial_y b&=&\frac{2i}{M}\langle\hat{\Psi}_1\partial_y^2\hat{\Psi}_1^{\dag}\rangle-\langle\hat{\Psi}_1^{\dag}\partial_y^2\hat{\Psi}_1\rangle\nonumber \\&=&\frac{2\sqrt{\pi}k}{M}(-ja_1)\langle\hat{\Psi}^{\dag}_{1}\hat{\Psi}_{1}\rangle, \\
\partial_y a&=& \frac{2i}{M}\langle\partial_y\hat{\Psi}_1\partial_x\hat{\Psi}_1^{\dag}\rangle+\langle\hat{\Psi}_1\partial_x\partial_y \hat{\Psi}_1^{\dag}\rangle-\langle\partial_y\hat{\Psi}_1^{\dag}\partial_x\hat{\Psi}_1\rangle-\langle\hat{\Psi}_1^{\dag}\partial_x\partial_y \hat{\Psi}_1\rangle-2i\langle\hat{\Psi}_1\hat{\Psi}_1^{\dag}\rangle\nonumber \\&&-2iy[\langle\hat{\Psi}_1\partial_y\hat{\Psi}_1^{\dag}\rangle-\langle\hat{\Psi}_1^{\dag}\partial_y\hat{\Psi}_1\rangle]\rangle\nonumber \\&=&-\frac{2a_1^2j^2}{M}\langle\hat{\Psi}^{\dag}_{1}\hat{\Psi}_{1}\rangle,\\
\partial_x b&=&\frac{2i}{M}\langle\partial_x\hat{\Psi}_1\partial_y\hat{\Psi}_1^{\dag}\rangle+\langle\hat{\Psi}_1\partial_x\partial_y\hat{\Psi}_1^{\dag}\rangle-
\langle\partial_x\hat{\Psi}_1^{\dag}\partial_y\hat{\Psi}_1\rangle-\langle\hat{\Psi}_1^{\dag}\partial_x\partial_y\hat{\Psi}_1\rangle\nonumber \\&=&\frac{8\pi^2k^2}{a_1^2M}\langle\hat{\Psi}^{\dag}_{1}\hat{\Psi}_{1}\rangle.
\end{eqnarray}


\begin{thebibliography}{99}

\bibitem{maldacena} J. M. Maldacena, The large-N limit of superconformal field theories and supergravity, Adv. Theor. Math. Phys. {\bf2} (1998) 231 [Int. J. Theor. Phys. 38 (1999) 1113] [hep-th/9711200].

\bibitem{gkp}S. S. Gubser, I. R. Klebanov and A. M. Polyakov, Gauge theory correlators from non-critical string theory, Phys. Lett. B {\bf428} (1998) 105 [hep-th/9802109].
\bibitem{witten}E. Witten, Anti-de Sitter space and holography, Adv. Theor. Math. Phys. {\bf2} (1998) 253 [hep-th/9802150].
\bibitem{kss} P. Kovtun, D. T. Son and A.O. Starinets, Holography and hydrodynamics: diffusion on stretched horizons, JHEP {\bf10} (2003) 064 [hep-th/0309213].
\bibitem{bl}A. Buchel and J.T. Liu, Universality of the shear viscosity in supergravity, Phys. Rev. Lett. {\bf93} (2004) 090602 [hep-th/0311175].

\bibitem{geshu} X. -H. Ge, Y. Matsuo, F. -W. Shu, S. -J. Sin and T. Tsukioka, Viscosity Bound, Causality Violation and
Instability with Stringy Correction and Charge, JHEP {\bf0810} (2008) 006 [hep-th/0808.2354]

\bibitem{hartnoll}S. A. Hartnoll, Lectures on holographic methods for condensed matter physics, Class. Quant. Grav. {\bf26} (2009) 224002 [arXiv:0903.3246 [hep-th]].
\bibitem{herzog}C. P. Herzog, Lectures on Holographic Superfluidity and Superconductivity, J. Phys. A {\bf42} (2009) 343001 [arXiv:0904.1975 [hep-th]].

\bibitem{cai1} R.-G. Cai, L. Li, L.-F. Li, R.-Q. Yang, Introduction to Holographic Superconductor Models, Sci China-Phys Mech Astron {\bf58} (2015) 060401 [arXiv:1502.00437
[hep-th]].

\bibitem{ling1} Y. Ling, C. Niu, J. Wu, Z. Xian, H. Zhang, Metal-insulator Transition by Holographic Charge Density Waves, Phys. Rev. Lett. {\bf113} (2014) 091602.
\bibitem{son}  D. T. Son, Toward an AdS/cold atoms correspondence: a geometric realization of the Schroedinger symmetry, Phys. Rev. D {\bf78} (2008) 046003.
\bibitem{klm}  S. Kachru, X. Liu and M. Mulligan, Gravity duals of Lifshitz-like Fixed Points, Phys. Rev. D {\bf78} (2008) 106005 [arXiv:0808.1725].
\bibitem{taylor}  M. Taylor, Non-relativistic holography, arXiv:0812.0530.

\bibitem{kiritsis}C. Charmousis, B. Gouteraux, B. S. Kim, E. Kiritsis and R. Meyer, Effective Holographic Theories for low-temperature condensed matter systems, JHEP {\bf1011} (2010) 151  [arXiv:1005.4690 [hep-th]];
B. Gouteraux and E. Kiritsis, Generalized Holographic Quantum Criticality at Finite Density, JHEP {\bf1112} (2011) 036 [arXiv:1107.2116 [hep-th]].
\bibitem{xdong} X. Dong, S. Harrison, S. Kachru, G. Torroba and H. Wang, Aspects of holography for theories with hyperscaling violation, JHEP {\bf06} (2012) 041 [arXiv:1201.1905].

\bibitem{panzhang} Q. Pan, S.-J. Zhang, Revisiting holographic superconductors with hyperscaling violation, arXiv:1510.09199.
\bibitem{lmx} Gu-Qiang Li, Jie-Xiong Mo, Xiao-Bao Xu, Entanglement temperature for black branes with hyperscaling violation, arXiv:1509.05985.
\bibitem{gv}  P. A. Gonzlez, Y. Vsquez, Scalar Perturbations of Nonlinear Charged Lifshitz Black Branes with Hyperscaling Violation, arXiv:1509.00802.
\bibitem{kym}  E. Kiritsis, Y. Matsuo, Hyperscaling-Violating Lifshitz hydrodynamics from black-holes, arXiv:1508.02494.
\bibitem{kofinas}G. Kofinas, Hyperscaling violating black holes in scalar-torsion theories,  Phys. Rev. D {\bf92} (2015) 084022 [arXiv:1507.07434].

\bibitem{dss}M. H. Dehghani, A. Sheykhi, S. E. Sadati, Thermodynamics of nonlinear charged Lifshitz black branes with hyperscaling violation, Phys. Rev. D {\bf91} (2015) 124073 [arXiv:1505.01134].
\bibitem{elp} D. Elander, R. Lawrance, M. Piai, Hyperscaling violation and Electroweak Symmetry Breaking, arXiv:1504.07949.
\bibitem{fengg}X.-H. Feng, W.-J. Geng, Non-Abelian (Hyperscaling Violating) Lifshitz Black Holes in General Dimensions,  Phys. Lett. B {\bf747} (2015) 395 [arXiv:1502.00863].
\bibitem{fl}Z.-Y. Fan, H. Lu, Electrically-Charged Lifshitz Spacetimes, and Hyperscaling Violations, arXiv:1501.05318.

\bibitem{kpww1}X.-M. Kuang, E. Papantonopoulos, B. Wang, J.-P.  Wu, Dynamically generated gap from holography in the charged black brane with hyperscaling violation, JHEP {\bf04} (2015) 137 [arXiv:1411.5627].
\bibitem{lsachdev} A. Lucas, S. Sachdev, Conductivity of weakly disordered strange metals: from conformal to hyperscaling-violating regimes, Nuclear Physics B {\bf892} (2015) 239 [arXiv:1411.3331].
\bibitem{kpww2}X.-M. Kuang, E. Papantonopoulos, B. Wang, J.-P.  Wu, Formation of Fermi surfaces and the appearance of liquid phases in holographic theories with hyperscaling violation, JHEP {\bf11} (2014) 086 [arXiv:1409.2945].

\bibitem{br}Pablo Bueno, Pedro F. Ramirez, Higher-curvature corrections to holographic entanglement entropy in geometries with hyperscaling violation, arXiv:1408.6380.
\bibitem{pdsr} Parijat Dey, Shibaji Roy, Interpolating solution from AdS$_5$ to hyperscaling violating Lifshitz space-time, arXiv:1406.5992.
\bibitem{horowitz1} G. T. Horowitz, J. E. Santos and D. Tong, Optical Conductivity with Holographic Lattices, JHEP {\bf1207} (2012) 168  [arXiv:1204.0519 [hep-th]].
\bibitem{horowitz2} G. T. Horowitz, J. E. Santos and D. Tong, Further Evidence for Lattice-Induced Scaling, JHEP {\bf1211} (2012) 102 [arXiv:1209.1098 [hep-th]].
\bibitem{dg1}  A. Donos and J.P. Gauntlett, Holographic helical superconductors, J. High Energy Phys. {\bf12} (2011) 091.
\bibitem{dg2}  A. Donos and J. P. Gauntlett, Black holes dual to helical current phases, Phys. Rev. D {\bf86} (2012) 064010.
\bibitem{donos}  A. Donos, Striped phases from holography, J. High Energy Phys. {\bf05} (2013) 059.
\bibitem{dh}   A. Donos and S. A. Hartnoll, Interaction-driven localization in holography, Nat. Phys. {\bf9} (2013) 649.
\bibitem{vegh}  D. Vegh, Holography without translational symmetry, arXiv:1301.0537.
\bibitem{davison}  R. A. Davison, Momentum relaxation in holographic massive gravity, Phys. Rev. D {\bf88} (2013) 086003 [arXiv:1306.5792].
\bibitem{btv}  M. Blake, D. Tong and D. Vegh, Holographic Lattices Give the Graviton an Effective Mass, Phys. Rev. Lett. {\bf112} (2014) 071602 [arXiv:1310.3832].
\bibitem{bt}  M. Blake and D. Tong, Universal Resistivity from Holographic Massive Gravity, Phys. Rev. D {\bf88} (2013) 106004 [arXiv:1308.4970].
\bibitem{dg3}  A. Donos and J. P. Gauntlett, Holographic Q-lattices, JHEP {\bf1404} (2014) 040 [arXiv:1311.3292].
\bibitem{aw} T. Andrade and B. Withers, A simple holographic model of momentum relaxation, JHEP {\bf1405} (2014) 101 [arXiv:1311.5157].
\bibitem{kr} E. Kiritsis, J. Ren, On Holographic Insulators and Supersolids, arXiv:1503.03481.
\bibitem{bhks} N. Bao, S. Harrison, S. Kachru, and S. Sachdev, Vortex lattices and crystalline geometrices, Phys. Rev. D {\bf88} (2013) 026002.
\bibitem{bh} N. Bao and S. Harrison, Crystalline Scaling Geometries frome Vortex Lattices, Phys. Rev. D {\bf88} (2013) 046009.
\bibitem{mozaffar} M. R. M. Mozaffar and A. Mollabashi, Crystalline Geometries from Fermionic Vortex Lattice, Phys. Rev. D {\bf89} (2014) 046007 [arXiv:1307.7397].
\bibitem{AMS} A. Allais, J. McGreevy, and S. J. Suh, A quantum electron star, Phys. Rev. Lett. {\bf108} (2012) 231602, [arXiv:1202.5308 [hep-th]].
\bibitem{hartnoll1}S. A. Hartnoll, A. Tavanfar, Electron stars for holographic mettalic criticality, Phys. Rev. D {\bf83} (2011) 046003 [arXiv: 1008.2828 [hep-th]].
\bibitem{hartnoll2}S. A. Hartnoll, D. M. Hofman and A. Tavanfar, Holographically smeared Fermi surface: Quantum oscillations and Luttinger count in electron star, Europhys. Lett. {\bf95} (2011) 31002, [arXiv:1011.2502[hep-th]].
\bibitem{dtong1} S. Bolognesi, J. N. Laia, D. Tong, K. Wong, A gapless hard wall: magnetic catalysis in bulk and boundary, JHEP {\bf07} (2012) 162.
\bibitem{maeda} K. Maeda, M. Natsuume, T. Okamura, Vortex lattice for a holographic superconductor,  Phys. Rev. D {\bf81} (2010) 026002.
\bibitem{shu} H. Guo, F. W. Shu, J. H. Chen, H. Li, Z. Yu, A holographic model of d-wave superconductor vortices with Lifshitz scaling, Int. J. Mod. Phys. D (to appear) [arXiv:1410.7020 [hep-th]].


\bibitem{HL} P. Ho\v{r}ava, Quantum Gravity at a Lifshitz Point, Phys. Rev. D {\bf79} (2009) 084008, [arXiv:0901.3775].
\bibitem{HL2} T. Griffin, P. Ho\v{r}ava, and C. M. Melby-Thompson, Lifshitz Gravity for Lifshitz Holography, Phys. Rev. Lett. {\bf110} (2013) 081602.
\bibitem{shu1} F.-W. Shu, K. Lin, A. Wang, and Q. Wu,  Lifshitz spacetimes, solitons and singularity, J. High Energy Phys. {\bf04} (2014) 056.
\bibitem{shu2} K. Lin, F.-W. Shu, A. Wang, and Q. Wu, Tidal deformation of a slowly rotating black hole, Phys. Rev. D {\bf91} (2015) 044003 [arXiv: 1403.3413[hep-th]].

\end{thebibliography}
\end{document}